\documentclass[preprint]{Interspeech}

\interspeechcameraready

\title{{Transformer Architectures for Respiratory Sound Analysis and Multimodal Diagnosis}}

\author[affiliation={1,2}]{Theodore}{Aptekarev$^\dagger$}
\author[affiliation={1}]{Vladimir}{Sokolovsky}
\author[affiliation={1}]{Gregory}{Furman}

\affiliation{}{Ben Gurion University of the Negev}{Israel}
\affiliation{}{HSE University}{Russia}

\email{aptekath@bgu.ac.il, sokolovv@bgu.ac.il, gregoryf@bgu.ac.il}

\makeatletter
\renewcommand{\affiliation}[3]{%
    \stepcounter{affcounter}%
    \def\affline{}%
    \def\affsep{}%
    \if\relax\detokenize{#1}\relax\else
        \edef\affline{#1}\def\affsep{, }%
    \fi
    \if\relax\detokenize{#2}\relax\else
        \edef\affline{\affline\affsep#2}\def\affsep{, }%
    \fi
    \if\relax\detokenize{#3}\relax\else
        \edef\affline{\affline\affsep#3}%
    \fi
    \edef\affiliationslist{\affiliationslist\affiliationsep$^{\theaffcounter}$\affline}%
    \renewcommand{\affiliationsep}{\vskip1pt}%
}
\makeatother

\usepackage{comment}
\usepackage{cite}
\usepackage{xurl}
\usepackage{rotating}
\usepackage[table]{xcolor}
\usepackage{makecell}
\usepackage{tabu}
\usepackage{tabularx}
\usepackage{ragged2e}
\usepackage{subcaption}
\usepackage{multicol}
\usepackage{multirow}
\usepackage{longtable}
\usepackage{underscore}
\makeatletter
\def\maxwidth{\ifdim\Gin@nat@width>\linewidth\linewidth\else\Gin@nat@width\fi}
\def\maxheight{\ifdim\Gin@nat@height>\textheight\textheight\else\Gin@nat@height\fi}
\makeatother
\setkeys{Gin}{width=\maxwidth,height=\maxheight,keepaspectratio}
\newcommand{\pandocbounded}[1]{#1}
\newcolumntype{Y}{>{\RaggedRight\arraybackslash}X}

\begin{document}

\maketitle
\renewcommand{\thefootnote}{$\dagger$}
\footnotetext{Corresponding author.}
\renewcommand{\thefootnote}{\arabic{footnote}}

\noindent\textit{Preprint.}\\

\begin{abstract}
Respiratory sound analysis is a crucial tool for screening asthma and
other pulmonary pathologies, yet traditional auscultation remains
subjective and experience-dependent \cite{gurung2011, murphy2004}. Our
prior research established a CNN baseline using DenseNet201, which
demonstrated high sensitivity in classifying respiratory sounds
\cite{aptekarev2023}. In this work, we (i) adapt the Audio Spectrogram
Transformer (AST) \cite{gong2021ast} for respiratory sound analysis and
(ii) evaluate a multimodal Vision-Language Model (VLM) that integrates
spectrograms with structured patient metadata.

AST is initialized from publicly available weights and fine-tuned on a
medical dataset containing hundreds of recordings per diagnosis. The VLM
experiment uses a compact Moondream-type model
\cite{korrapati2025moondream2} that processes spectrogram images
alongside a structured text prompt (sex, age, recording site) to output
a JSON-formatted diagnosis. Results indicate that AST achieves
approximately 97\% accuracy with an F1-score around 97\% and ROC AUC of
0.98 for asthma detection, significantly outperforming both the internal
CNN baseline and typical external benchmarks. The VLM reaches 86-87\%
accuracy, performing comparably to the CNN baseline while demonstrating
the capability to integrate clinical context into the inference process.
These results confirm the effectiveness of self-attention for acoustic
screening and highlight the potential of multimodal architectures for
holistic diagnostic tools.
\end{abstract}

\section{Introduction}\label{introduction}

Respiratory diseases represent a significant portion of global morbidity
and mortality. Bronchial asthma, in particular, affects up to 29\% of
the population in various countries \cite{mortimer2022, asher2021},
amounting to approximately 262 million people worldwide and causing over
460,000 deaths annually \cite{vos2019}. It consistently ranks among the
top 30 conditions affecting all age groups \cite{vos2019}. Traditional
diagnosis relies on a complex assessment including history taking,
spirometry, and auscultation. However, this process faces limitations:
manual auscultation is subjective and depends heavily on clinician
experience \cite{gurung2011, murphy2004}, functional tests are
difficult for young children, and access to specialized care is often
limited in remote regions. Studies suggest that up to 30\% of asthma
patients may receive an incorrect diagnosis when it is based solely on
symptoms \cite{tomita2019}.

\subsection{Related work and
baselines}\label{related-work-and-baselines}

Computerized analysis of respiratory sounds has evolved from early
digital signal processing to advanced machine learning, offering an
objective tool for screening and telemedicine
\cite{gurung2011, emmanouilidou2018, tomita2019}. Early systems focused
on distinguishing normal from pathological sounds using spectral
features \cite{urquhart1981, nissan1993, bahoura2006}. Later,
Convolutional Neural Networks (CNNs) have shown promise. Studies
comparing various CNN architectures (e.g., VGG, ResNet, DenseNet) for
respiratory pathology detection have reported accuracies ranging from
80\% to over 90\%, depending on the dataset and task
\cite{kim2021, palanisamy2020}.

However, comparing results across studies is challenging due to the lack
of standardization in datasets and recording protocols. As noted in
recent reviews \cite{ruchonnet2024}, public datasets vary significantly
in size, annotation quality (e.g., acoustic events vs.~clinical
diagnosis), and device heterogeneity. For instance, models trained on
cough-centric datasets like COUGHVID \cite{coughvid2021} cannot be
directly compared to those trained on auscultation breathing sounds.

To address this, we utilize a strictly controlled internal baseline.
Prior research on the specific dataset used in this study
\cite{aptekarev2023} demonstrated that DenseNet201 \cite{huang2016}
achieved high sensitivity in classifying respiratory sounds. By using
this established high-performing CNN as a reference point on the exact
same data splits, we ensure a fair evaluation of the proposed
Transformer architectures, isolating the impact of the model
architecture from data variations.

Recent reviews highlight a rapidly expanding landscape of respiratory
sound datasets and ML approaches. Xia et al.~(2022) surveyed 11 datasets
and reported that traditional ML methods typically achieve accuracy
around 75-82\%, while deep models using spectrograms reach about 85-90\%
in large datasets \cite{xia2022, ruchonnet2024}. The field also shifted
toward cough-centered datasets during the COVID-19 period (e.g.,
COUGHVID and Coswara), which accelerated publication volume but reduced
emphasis on breathing sounds
\cite{coughvid2021, coswara2023, santosh2022}. Existing public datasets
like ICBHI 2017 \cite{rocha_data2017} or HF\_Lung \cite{hf_lung2021}
often focus on detecting acoustic events (crackles, wheezes) rather than
nosological diagnosis, or are limited by specific recording equipment.

Clinical diagnosis is rarely based on a single signal source
\cite{gina2021}; it synthesizes patient history, demographics, and
symptoms alongside physical examination. In prior CNN pipelines, only
acoustic inputs were used and structured metadata were not integrated
into the classifier \cite{aptekarev2023}.

Recent studies have begun applying transformer architectures to
respiratory acoustics. Aljaddouh et al.~(2024) used Vision Transformer
models on respiratory spectrograms and reported accuracy around 91\% for
multi-class respiratory condition classification\cite{aljaddouh2024}.
For multimodal integration, Tang et al.~(2022) combined cough audio with
patient metadata using a hierarchical multimodal transformer for
COVID-19 detection \cite{tang2022}. Despite these advances,
transformer-based studies focused specifically on asthma screening from
breathing sounds remain limited, and the role of structured clinical
context in such settings has not been systematically evaluated.

\subsection{Objectives and
contribution}\label{objectives-and-contribution}

This study investigates the applicability of transformer architectures
for screening asthma from respiratory sounds. We address two objectives:
(i) adapt the Audio Spectrogram Transformer (AST) \cite{gong2021ast}
for respiratory sound classification to improve quality relative to the
CNN baseline, and (ii) evaluate a multimodal approach using a
Vision-Language Model (VLM) that combines spectrograms with structured
patient metadata (age, sex, recording site). This formulation mirrors
the clinical workflow by incorporating contextual patient information
into the inference process.

\section{Methods}\label{methods}

\subsection{Dataset context}\label{dataset-context}

The study utilized an anonymized database collected at the Regional
Children's Clinical Hospital of Perm Krai (Perm, Russia), as described
in previous work
\cite{aptekarev2023, furman2014, furman2020, gelman2022}. The protocol
was approved by the Ethics Committee of Perm State Medical University
and followed the Declaration of Helsinki \cite{wma2013}. The full
dataset comprises 1371 subjects (aged 0--47 years), including healthy
volunteers and patients with confirmed respiratory diagnoses such as
asthma (diagnosed according to GINA guidelines \cite{gina2021}). The
dataset is institutional property of Perm State Medical University and
is available on request \cite{aptekarev2023}.

Inclusion criteria required a verified diagnosis of a respiratory
disease (for asthma, with clinical staging), absence of acute
respiratory infection at the time of recording, and informed consent.
Exclusion criteria included other acute respiratory diseases,
decompensated chronic conditions, and refusal to participate
\cite{aptekarev2023}. Each recording is linked to clinical annotations
and structured metadata (sex, age, recording point, diagnosis, record
date, and quality flags), enabling the multimodal experiments described
below. A more detailed description of the data collection protocol and
metadata schema is available in \cite{aptekarev2023}.

Recordings were performed during quiet breathing at four anatomical
points (mouth, trachea, right second intercostal space, and right
paravertebral area). Each recording lasted several breathing cycles,
averaging about 25 seconds (minimum 16 seconds). Data were collected
using various hardware ranging from specialized computerized systems
with external microphones and electronic stethoscopes to mobile phones.
These systems ensure amplitude-frequency linearity in the 100--3000 Hz
range and have been validated in prior technical and clinical studies
\cite{furman2014, furman2020, gelman2022, reichert2009}. Sampling rates
varied from 22 kHz to 96 kHz. Original files were stored as WAV, MP3, or
M4A and were harmonized to WAV during preprocessing. The dataset
includes clinician-assigned quality labels and automated technical
defect flags, which were used to filter out defective recordings during
preprocessing \cite{aptekarev2023}.

The 1,613 recordings include multiple recordings per subject and
multiple recording points per patient. Tracheal recordings dominate
(about 60\%), followed by chest, back, and mouth. Recording sources
include specialized systems (878 recordings), mobile phones (345),
remote web collection (85), and standard computers (63), with 242
recordings lacking source metadata \cite{aptekarev2023}. The dataset is
single-center, which reduces device variability but limits external
validity.

Quality control combines clinician labels (\texttt{record\_quality})
with automated flags: \texttt{technical\_defect} marks recordings
shorter than 14 s or with decoding errors, and
\texttt{amplitude\_defect} marks excessive clipping (above
\textasciitilde2\% of samples). Only recordings labeled as good/average
and without defect flags are retained for model training. After
filtering, audio is trimmed to remove transient artifacts and normalized
in amplitude before segmentation into clips \cite{aptekarev2023}.

\begin{table}[t]
\centering
\caption{Dataset summary}
\label{tab:dataset}
\begin{tabularx}{\columnwidth}{@{}YY@{}}
\toprule
Metric & Details \\
\midrule
Total patients and volunteers & 1,613 \\
Sex distribution & F (542), M (1,071) \\
Age distribution &  0--47 years \\
Conditions & Asthma (1,113), \newline Healthy (133), \newline Other pathologies (367) \\
Recording points & Trachea (\textasciitilde60\%), \newline Chest, \newline Back, \newline Mouth \\
\bottomrule
\end{tabularx}
\end{table}

Table 1 summarizes the dataset characteristics. Other pathologies in the
dataset include cystic fibrosis, recurrent obstructive bronchitis,
pneumonia, and bronchopulmonary dysplasia \cite{aptekarev2023}.

We report the ``Asthma vs Not Asthma'' binary classification results
derived from this dataset. This target was selected to determine whether
the model can distinguish asthma from other pathologies (including
healthy subjects and other respiratory diseases), a clinically critical
task given that different conditions may manifest with similar audio
features (e.g., wheezing) in standard auscultation.

To ensure a fair comparison, all models in this study (DenseNet, AST,
and VLM) were trained and evaluated on the same dataset. While the
specific random splits may have varied between the initial DenseNet
experiments and the subsequent Transformer evaluations, the underlying
data pool, preprocessing protocols, and class definitions remained
consistent to minimize variance in the results.

\subsection{Audio spectrogram transformer
(AST)}\label{audio-spectrogram-transformer-ast}

Audio Spectrogram Transformer (AST) \cite{gong2021ast} is a transformer
model designed for audio spectrograms. The input 2D spectrogram is split
into small patches, linearized, and passed through a transformer encoder
with positional embeddings. A special class token is added and its
output is used for classification, analogous to BERT
\cite{devlin2018bert} and ViT \cite{dosovitskiy2020vit}.

AST uses transfer learning from a vision transformer pretrained on
ImageNet \cite{deng2009}. In the original implementation, AST weights
are initialized from ViT and then fine-tuned on AudioSet
\cite{gemmeke2017audioset}, a large audio dataset. The experiment in
this study uses publicly available weights produced using this
procedure. AST is reported to achieve about 95.6\% accuracy on ESC-50
\cite{piczak2015esc50}, which motivated its selection.

\subsubsection{Fine-tuning AST on medical
data}\label{fine-tuning-ast-on-medical-data}

AST is larger than DenseNet but still moderate in size for transformers
(about 87 million parameters in the reference implementation), while the
medical dataset is limited (hundreds of recordings per class, split into
several thousand clips). To reduce overfitting, the training used
increased weight decay, gradient clipping, and early stopping when
metrics stagnated. Extensive audio augmentation (noise addition, time
shifts, pitch changes) was not applied in order to preserve clinical
validity. In medical audio, such transformations can distort clinically
significant markers (e.g., wheezes) or erase subtle acoustic patterns. A
small learning rate (around \(10^{-5}\) and below) with warmup was used.

\subsection{Multimodal vision-language model
(VLM)}\label{multimodal-vision-language-model-vlm}

A Moondream-type VLM was selected for the multimodal experiment. It
expects an RGB image and a text prompt.

\subsubsection{VLM data preparation and
prompts}\label{vlm-data-preparation-and-prompts}

Respiratory audio was converted to mel-spectrogram images using the
approach described in \cite{aptekarev2023}. Mel-spectra were computed
with a 0-8 kHz frequency range, 128 mel coefficients, and a 20-25 ms
analysis window. To build a three-channel input, each audio fragment was
converted into an RGB image where channels correspond to different
window sizes: short (25 ms), medium (100 ms), and long (175 ms). Unlike
the baseline DenseNet201 \cite{huang2016} pipeline that used 3-channel
mel-spectrograms normalized to {[}0,1{]}, the VLM uses standard RGB
images in {[}0,255{]}. Figure 1 shows an example TIFF spectrogram used
as input.

\begin{figure}
\centering
\pandocbounded{\includegraphics[keepaspectratio,alt={Example TIFF spectrogram input}]{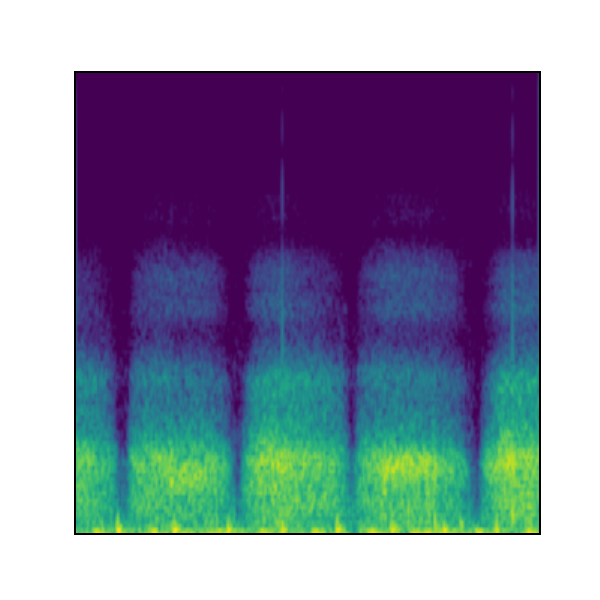}}
\caption{Example TIFF spectrogram input}
\end{figure}

In addition to the visual input, the model receives structured text
metadata (sex, age, recording site such as mouth, trachea, or lung area)
and a task instruction. The structured prompt used in the experiments is
shown in Figure 2. Key names are in English to keep model inputs and
JSON parsing consistent.

\begin{figure*}
\centering
\pandocbounded{\includegraphics[keepaspectratio,alt={VLM inference pipeline from spectrogram and prompt to JSON output}]{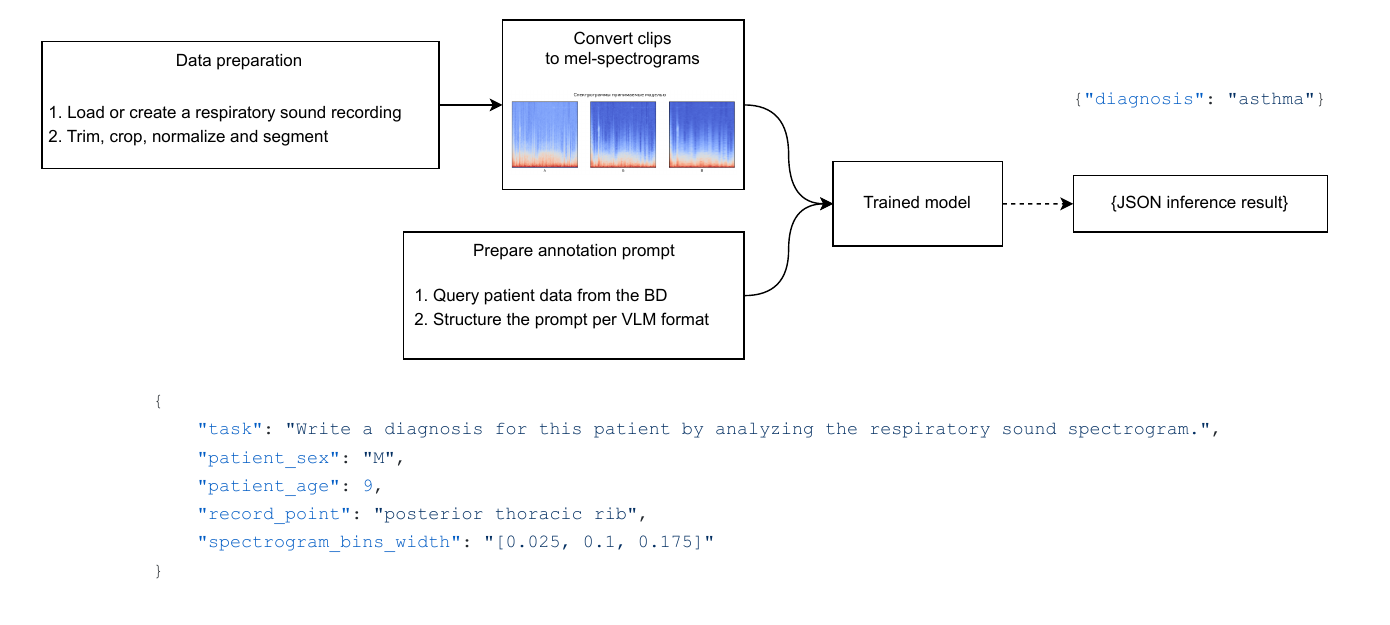}}
\caption{VLM inference pipeline from spectrogram and prompt to JSON
output}\label{fig:vlm-pipeline}
\end{figure*}

Two prompt formats were tested: natural language and JSON-like
structured text. Pilot runs on a balanced subset showed no measurable
difference in classification accuracy, so the structured format was
chosen because it simplifies the automatic analysis of model outputs
(Figure 2).

The VLM experiments used a balanced pool of 200 patients (100 per
class). The split into training and test sets was stratified to preserve
class balance and to avoid clip overlap between training and evaluation.

\subsubsection{VLM fine-tuning and
inference}\label{vlm-fine-tuning-and-inference}

The VLM was adapted by fine-tuning on the target classification task.
Given that open weights were pretrained on large image (e.g., ImageNet
\cite{deng2009}) and text corpus, transfer learning was used to reduce
computational cost. For the 1.9B-parameter Moondream2 model
\cite{korrapati2025moondream2}, fine-tuning used Low-Rank Adaptation
(LoRA).

Adapters were added to key modules of the visual encoder and the
language model, while core weights remained frozen to preserve
pretrained knowledge. For the Moondream2 architecture (Phi
\cite{li2023phi15} / ViT), this included \texttt{Wqkv} and
\texttt{out\_proj} in the text model, \texttt{fc1} and \texttt{fc2} in
intermediate MLP layers, and the \texttt{proj} layer that maps visual
features into the language embedding space. Adapter parameters accounted
for about 3.8\% of the total parameter count, keeping the trainable
footprint small relative to full fine-tuning. The final classifier head
and embeddings were unfrozen and trained along with the adapters.

Training ran for 3 epochs (125 batch iterations) over several thousand
clips, with a batch size around 30 and the Adam optimizer in 8-bit mode
to reduce memory usage. A cosine learning rate schedule with a short
warmup was used. During evaluation, the model produced a JSON-formatted
diagnosis (e.g., \texttt{\{"diagnosis":\ "asthma"\}}); outputs were
normalized for JSON validity and compared with reference labels.

\section{Results}\label{results}

\subsection{Input duration and AST
outcomes}\label{input-duration-and-ast-outcomes}

AST was trained successfully with both 10-second and 5-second fragments.
The 5-second setting did not reduce quality relative to 10-second clips.
Shorter clips increase the number of training samples, while longer
clips provide more context but reduce the number of examples. Given
these considerations, 5-second clips were used for the final evaluation.
The high classification accuracy achieved with this duration confirms
that 5-second segments capture sufficient acoustic information for
reliable diagnosis while maximizing the training data available from
limited recordings.

\subsection{AST classification
results}\label{ast-classification-results}

For the primary ``Asthma vs Not Asthma'' task, AST achieved
approximately 97\% accuracy with an F1-score around 97\% and ROC AUC of
0.98. These results demonstrate a significant improvement over the
baseline CNN models. In broader screening settings, AST also maintained
high performance, reaching similar accuracy levels (\textasciitilde97\%)
for detecting ``Any Pathology'' versus healthy controls.

\subsection{VLM classification
results}\label{vlm-classification-results}

After fine-tuning, the multimodal VLM achieved about 86-87\% accuracy
for the ``Asthma vs Not Asthma'' task, with best results of Accuracy =
86.5\% and F1-score = 87.7\%. The Youden index was around 0.73, which is
comparable to the DenseNet baseline (Youden index \textasciitilde0.74)
reported in prior work \cite{aptekarev2023}. A longer training variant
(6 epochs or 250 batch iterations) achieved about 85.5\% accuracy with
higher precision (Precision = 90.8\%, F1 = 84.5\%), indicating a
trade-off between sensitivity and false-positive rates.

\subsubsection{Ablation study: the role of
metadata}\label{ablation-study-the-role-of-metadata}

Selective prompt ablations showed that input text was required for
stable inference in this multimodal architecture. Removing the technical
text block that specifies spectrogram parameters (e.g., bin widths) led
to a drop in performance (Accuracy = 0.670, Youden Index = 0.340),
though the model retained partial predictive capacity.

In this VLM setup, removing the demographic block (patient age, sex,
recording point) caused a collapse of the classifier, producing 100\%
false negatives for asthma. In this configuration, the model relied on
the text conditioning to reach a stable decision boundary.

\subsection{Comparative analysis across
architectures}\label{comparative-analysis-across-architectures}

DenseNet201 from previous work \cite{aptekarev2023} achieved about 87\%
accuracy on a balanced test set for ``Asthma vs Not Asthma'', with
sensitivity around 93\% and specificity about 82-86\%. These values
serve as the reference point. Figure 3 compares the VLM and DenseNet
baselines for the same ``Asthma vs Not Asthma'' task.

\begin{figure}
\centering
\pandocbounded{\includegraphics[keepaspectratio,alt={Model performance comparison for Asthma vs Not Asthma}]{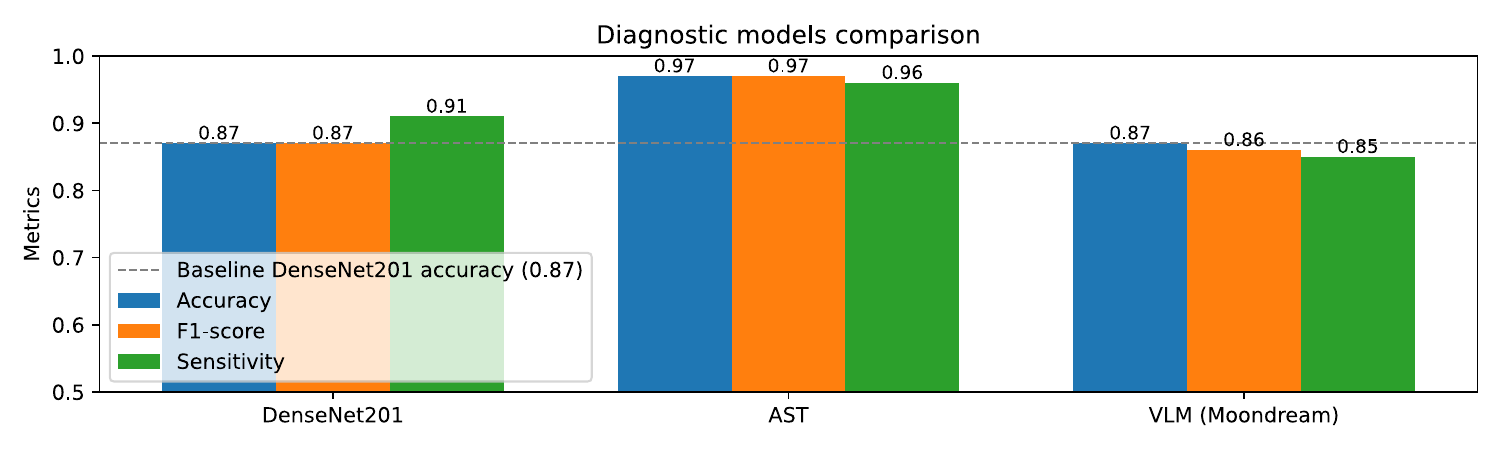}}
\caption{Model performance comparison for Asthma vs Not Asthma}
\end{figure}

The VLM achieved quality comparable to DenseNet on the ``Asthma vs Not
Asthma'' detection task (86-87\% accuracy, F1 around 0.87, Youden about
0.73). The VLM pipeline also produced a structured JSON diagnosis and
used structured patient metadata alongside spectrogram inputs.

\subsubsection{External context}\label{external-context}

While direct comparison with external studies is complicated by dataset
heterogeneity, our results align with and potentially exceed current
benchmarks in the field. For instance, Kim et al.~(2021) reported CNN
accuracies around 85.7\% for respiratory anomaly detection, and
systematic reviews found deep learning approaches typically reaching
85-90\% on large datasets \cite{xia2022}. Our AST model's 97\% accuracy
on the ``Asthma vs Not asthma'' task suggests that transformer
architectures may offer a significant performance leap over these
traditional CNN baselines when applied to high-quality, standardized
data. However, the strict internal comparison with DenseNet remains the
most scientifically valid measure of improvement within this study.

\section{Discussion}\label{discussion}

\subsection{Interpretability and
analysis}\label{interpretability-and-analysis}

\begin{figure*}
\centering
\pandocbounded{\includegraphics[keepaspectratio,alt={Interpretability via visualization: Grad-CAM heatmap overlay on a spectrogram}]{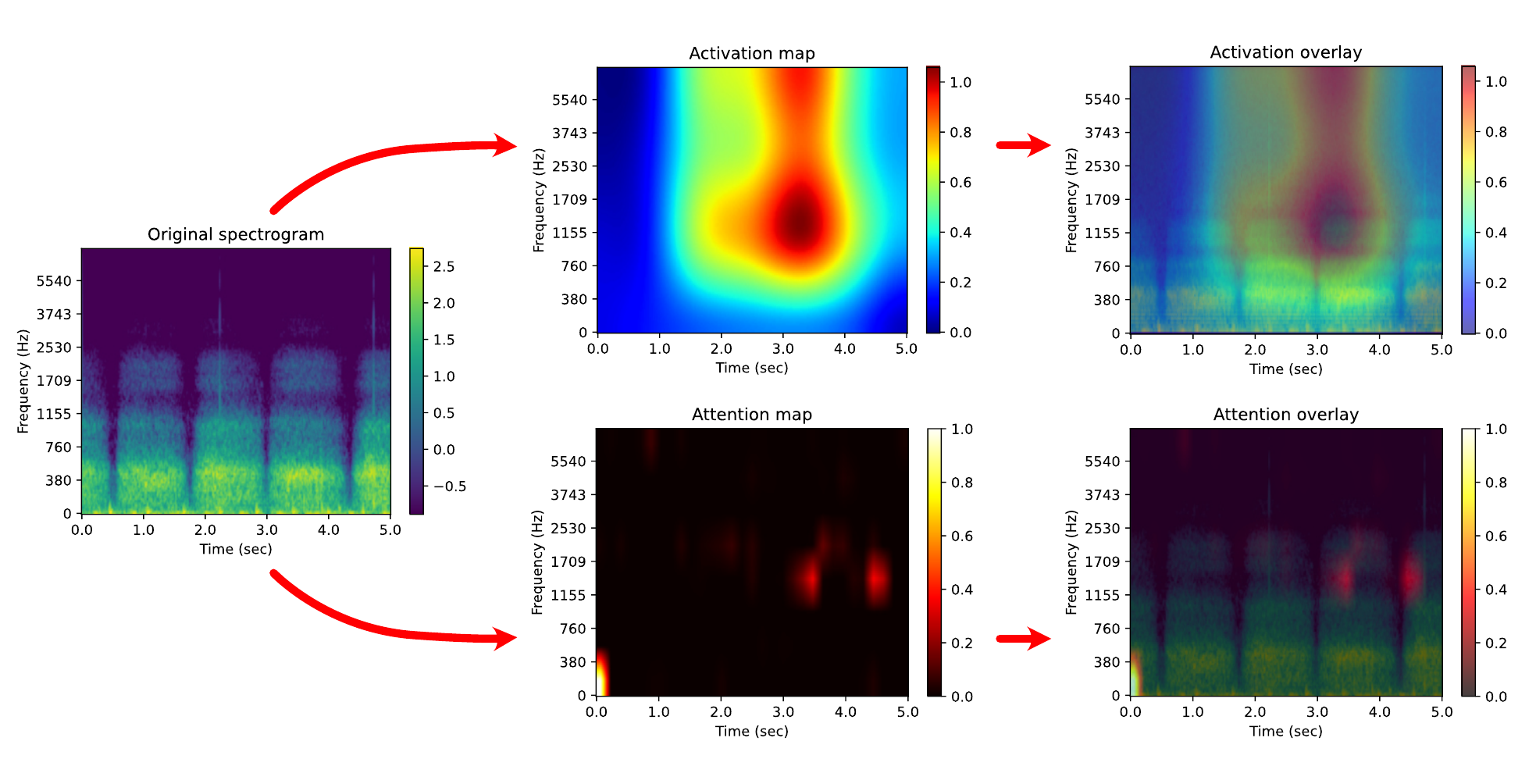}}
\caption{Interpretability via visualization: Grad-CAM heatmap overlay on
a spectrogram}\label{fig:interpretability}
\end{figure*}

Interpretability is an important consideration for clinical
applications. For DenseNet and other CNNs, Grad-CAM
\cite{selvaraju2017gradcam} can produce heatmaps that highlight
spectrogram regions most responsible for the predicted class. These
visualizations can indicate frequency bands and time intervals that
drive a diagnosis (Figure 4).

For AST, Grad-CAM is not directly applicable due to the absence of
convolutional layers. Instead, attention map analysis can be used. AST
splits the spectrogram into patches; by averaging attention matrices
across heads, a two-dimensional attention map can be reconstructed. This
can show which time-frequency regions contribute most to the class
token.

For VLMs, interpretability can be enhanced by generating not only a
diagnosis label but also a textual rationale (e.g., a description of
detected wheezes). Implementing this requires additional training data
with detailed expert annotations (e.g., time-aligned descriptions of
specific acoustic events or data derived from other clinical
modalities). This shifts the task toward image captioning for
spectrograms.

\subsection{Computational efficiency and
deployment}\label{computational-efficiency-and-deployment}

All evaluated models operate near real time. DenseNet (about 20M
parameters) required fractions of a second on GPU and roughly 2-5
seconds on CPU. In our measurements, AST (about 87M parameters) also
produced results within a few seconds on CPU, with sub-second inference
on GPU. The VLM required about 1.8 seconds per 5-second clip on CPU and
less than 0.5 seconds on GPU.

\section{Conclusion}\label{conclusion}

This study evaluated the application of Transformer-based architectures
for the automated screening of asthma from respiratory sounds. By
training and evaluating models on a controlled dataset, we demonstrated
that the Audio Spectrogram Transformer (AST) significantly outperforms
the established CNN baseline, achieving approximately 97\% accuracy and
F1-score for the ``Asthma vs Not Asthma'' task. These results, which
exceed typical state-of-the-art benchmarks for similar tasks (85--90\%),
confirm that the self-attention mechanism is highly effective at
capturing the complex, non-local acoustic patterns characteristic of
asthmatic breathing, even within short 5-second clips.

Furthermore, we successfully validated a multimodal approach using a
Vision-Language Model (VLM). While its quantitative performance
(86--87\% accuracy) was comparable to the CNN baseline rather than
superior to it, the VLM demonstrated the capability to integrate
structured patient metadata (demographics, recording site) directly into
the inference process. This mimics the clinical workflow where diagnosis
relies on context, not just signal, and produces structured,
interpretable JSON outputs suitable for integration into clinical
decision support systems.

Our findings suggest that while pure audio transformers like AST
currently offer the highest diagnostic accuracy for screening,
multimodal architectures pave the way for more holistic diagnostic
tools.

\bibliographystyle{IEEEtran}
\bibliography{mybib}

\end{document}